\documentclass[aps,prl,twocolumn,superscriptaddress,showpacs,floatfix,longbibliography]{revtex4-1}
\usepackage{amsmath,amssymb}

\usepackage[utf8]{inputenc}
\usepackage[T1]{fontenc}
\usepackage{xcolor}

\IfFileExists{newtxtext.sty}
   {\usepackage{newtxtext,newtxmath}}
   {\IfFileExists{stix.sty}
      {\usepackage{stix}}
      {\IfFileExists{mathptmx.sty}
      {\usepackage{mathptmx}}{} } }

\usepackage{textcomp}

\usepackage{bm}

\IfFileExists{siunitx.sty}{\usepackage{booktabs,siunitx}}{}

\pdfoutput=1
\usepackage{color}
\definecolor{LinkColor}{rgb}{0.256,0.439,0.588}
\usepackage{hyperref}
\hypersetup{
   pdfauthor={abc},
   pdftitle={paper},
   colorlinks=true,
   citecolor=blue,
   linkcolor=blue,
   urlcolor=blue
}

\usepackage{pifont}

\usepackage[pdftex]{graphicx}
\DeclareGraphicsExtensions{.pdf,.jpeg,.png}
\usepackage{epstopdf}

\begin{document}
	\bibliographystyle{apsrev4-1}
	\title{Hexagon Ising-Kondo lattice: An implication for intrinsic antiferromagnetic topological insulator}
	
	\author{Wei-Wei Yang}
	\affiliation{School of Physical Science and Technology $\&$ Key Laboratory for Magnetism and Magnetic Materials of the MoE, Lanzhou University, Lanzhou 730000, People Republic of China} %

	\author{Yin Zhong}
	\email{zhongy@lzu.edu.cn}
	\affiliation{School of Physical Science and Technology $\&$ Key Laboratory for Magnetism and Magnetic Materials of the MoE, Lanzhou University, Lanzhou 730000, People Republic of China} %

	\author{Hong-Gang Luo}
	\affiliation{School of Physical Science and Technology $\&$ Key Laboratory for Magnetism and Magnetic Materials of the MoE, Lanzhou University, Lanzhou 730000, People Republic of China} %
	\affiliation{Beijing Computational Science Research Center, Beijing 100084, China}%
	
	\begin{abstract}
		Recently, the MnBi$_2$Te$_4$ material has been proposed as the first intrinsic antiferromagnetic topological insulator (AFMTI), where the interplay between magnetism and topology induces several fascinating topological phases, such as the quantum anomalous Hall effect, Majorana fermions, and axion electrodynamics. %Despite of extensive theoretical and experimental studies, due to the strong correlation rarely microscopic model could deduce an exactly ground state of antiferromagneitc topological insulator.
		%Since Li et al.~\cite{PhysRevB.100.134438} have revealed that the magnetic anisotropy of 2-dimensional MnBi$_2$Te$_4$ originates from single-ion anisotropy,
%
However, an exactly solvable model being capable to capture the essential physics of the interplay between magnetism and topology is still absent. Here, inspired by the the Ising-like nature [B. Li \textit{et al.} Phys. Rev. Lett. \textbf{124}, 167204 (2020)] and the topological property of MnBi$_2$Te$_4$, we propose a topological Ising-Kondo lattice (TIKL) model to study its ground state property in an analytical way at zero temperature. The resultant phase diagram includes rich topological and magnetic states, which emerge in the model proposed in a natural and consistent way for the intrinsic magnetic topological insulator. With Monte Carlo simulation, we extend the AFMTI ground state to finite temperature. It reveals that topological properties do sustain at high temperature, which even can be restored by elevated temperature at suitable correlation strength. The results demonstrate that TIKL may offer an insight for future experimental research, with which magnetism and transport properties could be fine tuned to achieve more stable and exotic magnetic topological quantum states.

	\end{abstract}
	
	\date{\today}
	
	\maketitle
	
	%\tableofcontents

	\textit{Introduction.}$-$The quantized anomalous Hall (QAH) effect in magnetic topological insulator (TI) was demonstrated by Zhang \textit{et al.} a decade ago \cite{RevModPhys.83.1057}, where the ferromagnetic (FM) order that breaks the time-reversal symmetry is a crucial criterion for realizing the QAH state. For a long time, the QAH effect was only observed at sufficiently low temperature ($\sim1K$) in some weakly correlated materials \cite{Chang167,Yu61,Wang_2015,PhysRevLett.90.206601,PhysRevLett.105.057401,PhysRevB.94.085411,Liue1600167}, \textit{e.g.}, (Bi$_x$Sb$_{1-x}$)$_2$Te$_3$ thin films, where the topology sensitively depends on magnetic doping. %and some magnetic topological heterostructures.%引用science advances 的JIAHeng Li的引用25-31.
	When considerable electron correlation is taken into account, a growing variety of new regimes show up, such as the quantum spin liquid state \cite{Pesin} and a dynamical axion field \cite{Li2010}.
	By suppressing the effective bandwidth,
	correlations enhance the effect of spin-orbit interaction and lead to enhanced stability of the TI state with interactions, with which robust FM materials can spontaneously exhibit the QAH state without any external magnetic field. So far, even without introducing spin-scattering centres, the robust long-range ferromagnetism in topological materials has also been intensively realized in experiments \cite{Chang2015,cuizu,Katmis2016}.
	
	%However. due to the challenging in synthetizing antiferromagnetic (AFM) materials, the topological materials with AFM has rarely been observed.
	Compared with the above-mentioned TIs realized by magnetic doping or by proximity to a FM insulator with molecular beam expitaxy, the intrinsic magnetic TI is a more ideal and clean platform to display robust topological behaviors \cite{Smejkal2018}. However, due to complex magnetic structure, the study of antiferromagnetic (AFM) topological state is greatly hindered in experiment, despite tremendous efforts \cite{PhysRevB.90.041109}. Recently, the van der waals layered MnBi$_2$Te$_4$ family materials \cite{PhysRevLett.122.206401,Lieaaw5685,Otrokov2019} have been observed to be the antiferromagnetic topological insulator (AFMTI), in which the topology and intrinsic magnetism are firstly combined together. With the breaking time-reversal $\Theta$, breaking primitive-lattice translational symmetry $T_{1/2}$ but a preserving $S=\Theta T_{1/2}$, the MnBi$_2$Te$_4$ system leads to $Z_2$ topological classification.
	
	In MnBi$_2$Te$_4$ family materials, the interplay between topology and magnetism intriguingly generates numerous exotic topological quantum states, including the magnetic Weyl semimetal, nodal line system, AFMTI, or Chern insulator \cite{Chowdhury2019}. Although the conception about AFMTI has been proposed theoretically early in 2010 \cite{PhysRevB.81.245209}, its successful synthesis was fulfilled almost a decade later, which highlights the significance of an exactly analytical study about the intrinsic magnetic TI. So far, the relevant theoretical studies were limited to three ways: numerical simulations, such as first-principle calculation \cite{PhysRevB.91.235128}, Monte Carlo simulation \cite{PhysRevLett.106.100403} and variational cluster approximation \cite{PhysRevB.87.195133}; minimal model with artificially introduced AFM order \cite{Liue1600167,PhysRevB.81.245209,PhysRevB.89.035410}; and some approximation methods, including mean-field approximation \cite{PhysRevB.79.245331,PhysRevB.82.075125,PhysRevB.82.161302,PhysRevLett.100.156401} and the slave-particle method \cite{Pesin,PhysRevB.82.075106}. However, the exactly analytical study about the microscopic mechanism of the interplay between magnetic correlations and spin-orbit coupling (SOC) has not been carried out yet. Although some relevant theoretical studies have depicted the competition between magnetism and topology \cite{PhysRevB.79.245331,PhysRevB.82.075106,PhysRevB.82.075125,PhysRevLett.100.156401}, their coexistence is rarely realized.

%Therefore, an exactly analytical solvable model which could describe the behavior of intrinsic AFM topological materials is highly desired.
	
%	In this paper, we propose a microscopic model, i.e., the topological Ising-Kondo lattice (TIKL), to realize the $N\acute{e}el$ antiferromagnetic topological ground state. In the TIKL the magnetic topological states are generated in a similar mechanism with MnBi$_2$Te$_4$ family materials, where Mn atoms introduce long-range magnetic order and the Bi-Te layers generate topological states \cite{Lieaaw5685}, corresponding to the local and itinerant electrons, respectively. With anistropic Kondo interaction, the so-called TIKL model is exactly solvable at zero temperature. We study the interplay between SOC and Kondo interaction on honeycomb lattice, in which the analytical ground state and finite-temperature phase diagram calculated by Monte Carlo simulation are given. In addition, by analyzing possible low-lying excitations, we interpret the emergement of several exotic phenomenon at finite-temperatures.  Here, our results demonstrate that the notrivial magnetic state is not limited to low temperature.
%	% which sheds light on the realization of AFMTI in heavy fermion systems.

In this paper, motivated by the strong desire but incompatible lack of analytical research about intrinsic magnetic TI, we propose an exactly solvable microscopic model based on the topological properties and Ising-like nature \cite{PhysRevLett.124.167204} of MnBi$_2$Te$_4$. In MnBi$_2$Te$_4$, Mn atoms introduce long-range magnetic order and the Bi-Te layers generate topological states. Therefore, in the proposed topological Ising-Kondo lattice (TIKL) we use local electrons to mimic the Mn atoms, and the itinerant electrons to mimic Bi-Te layers, respectively. Thus some interesting behaviors of Mn(Sb$_x$Bi$_{(1-x)}$)$_2$Te$_4$ materials could be understood by the analytical study of exactly solvable TIKL, which successfully describes the $x$-dependent topological phase transition \cite{chen2019intrinsic} in Mn(Sb$_x$Bi$_{(1-x)}$)$_2$Te$_4$ family materials with the next-nearest neighbor (NNN) hopping dependent topological phase transition. With TIKL we systematically study the interplay between magnetism and topology. The accurate ground state phase diagram including rich magnetic topological states is mapped to offer some understanding about the essential physics of intrinsic magnetic TI. Besides, we show that the AFMTI ground state can be extended to finite temperature with Monte Carlo simulation. We also elaborate the finite temperature phase diagram, which demonstrates that topology survives at high temperature. At suitable correlation strength, elevated temperature could even drive the restoration of topological properties. Therefore, we analyse some possible low-lying excitations to explicate these exotic behaviors. Our findings demonstrate that the competition and coexistence of magnetism and topology can be understood within the context of TIKL, which may motivate further exploration of intrinsic magnetic TI.

\begin{figure}[htp!]
	\centering
	\includegraphics[width=3in]{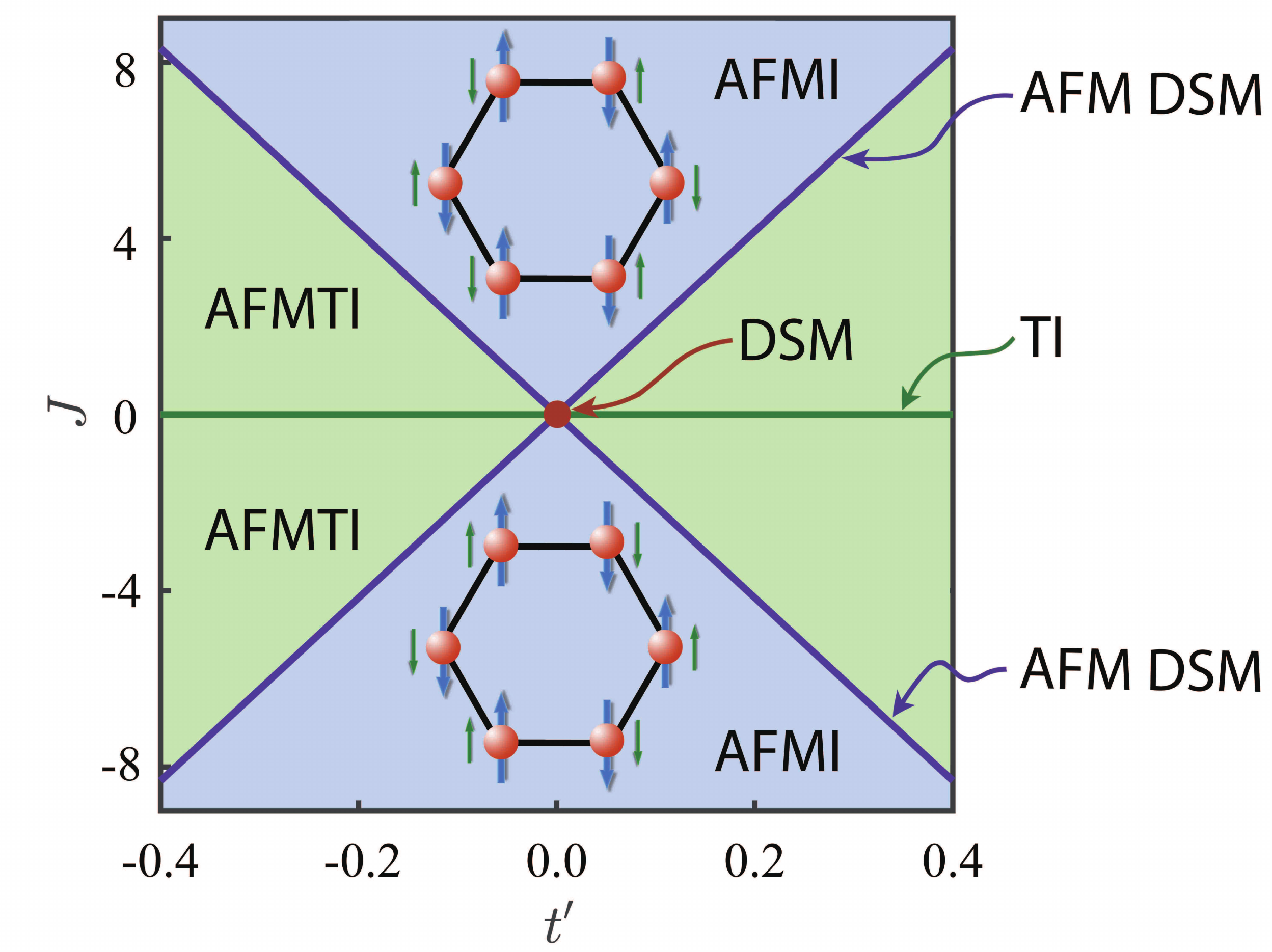}
	\caption{ The ground-state phase diagram of topological Ising-Kondo lattice (TIKL) model in the $J-t'$ plane. Rich quantum states include trivial antiferromagnetic insulator (AFMI), paramagnetic topological insulator (TI), antiferromagnetic topological insulator (AFMTI), paramagnetic Dirac semimetal (DSM) and antiferromagnetic Dirac semimetal (AFM DSM). %TI is exhibit at $J=0$, $t'\neq0$. AFM DSM is exhibit at the boundary of AFMTI-AFMI transition.
The inset demonstrates the antiferromagnetic (ferromagnetic) coupling between localized and itinerant electrons at $J>0$ $(J<0)$ region.}
	\label{fig:0T}
\end{figure}

	\textit{Model.}$-$Our starting point is the TIKL model on honeycomb lattice, which possesses two kinds of electrons, \textit{i.e.}, itinerant $c$-electrons and localized $f$-electrons. The localized $f$-electrons do not hop due to their infinite mass and only interact with $c$-electrons via longitudinal Kondo exchange. The $c$-electrons are modelled by the spinful Kane-Mele model and the whole Hamiltonian reads:
	\begin{equation}
	\hat{H}=-t\sum_{\langle i,j\rangle\sigma}\hat{c}_{i\sigma}^{\dag}\hat{c}_{j\sigma}
	-t'\sum_{\langle\langle ij\rangle\rangle\sigma}\sigma e^{i\phi_{ij}}\hat{c}_{i\sigma}^{\dag}\hat{c}_{j\sigma}+
	\frac{J}{2}\sum_{j\sigma}S_{j}^z\sigma \hat{c}_{j\sigma}^{\dag}\hat{c}_{j\sigma},
	\label{eq:model}
	\end{equation}
	where $\hat{c}_{j\sigma}^{\dag}(\hat{c}_{j\sigma})$ is $c$-electron's creation (annihilation) operator with spin $\sigma=\uparrow,\downarrow$ at site $j$. $S_{j}^{z}$ denotes the localized moment of $f$-electron. $t$-term denotes the nearest-neighbor (NN) hopping, and is used as the unit of energy ($t=1$). $t'$-term denotes the NNN hopping, which corresponds to the SOC strength. $\phi_{ij}=\pm\frac{\pi}{2}$, where the positive/negative phase corresponds to the anti-clockwise/clockwise hopping. $J$ is the longitudinal Kondo coupling. As shown in Fig.~\ref{fig:0T}, for $J>0$ ($J<0$) the coupling between localized and itinerant electrons is AFM (FM), while the AFM configuration of localized moments is maintained. Considering the symmetry of the phase diagram, we focus on the $J\geqslant0$ and $t'\geqslant0$ in the rest of our work.

	Choosing eigenstates of $S_{j}^{z}$ as basis, Eq.~(\ref{eq:model}) reduces into an effective free fermion model \cite{PhysRevB.100.045148}
	\begin{equation}
	\hat{H}=-t\sum_{\langle i,j\rangle\sigma}\hat{c}_{i\sigma}^{\dag}\hat{c}_{j\sigma}
	-t'\sum_{\langle\langle ij\rangle\rangle\sigma}\sigma e^{i\phi_{ij}}\hat{c}_{i\sigma}^{\dag}\hat{c}_{j\sigma}+
	\sum_{j\sigma}\frac{J\sigma}{4} q_{j} \hat{c}_{j\sigma}^{\dag}\hat{c}_{j\sigma}
	\label{eq:model2}
	\end{equation}
	under fixed background $\{q_{j}\}$, where $q_{j}=\pm1$ and $S_{j}^{z}|q_{j}\rangle=\frac{q_{j}}{2}|q_{j}\rangle$. Eq.~(\ref{eq:model2}) has the U(1) symmetry which reflects the charge conservation, and the $Z_2$ symmetry reflecting invariance under spin-$\uparrow$ $\rightarrow$ spin-$\downarrow$. At the half-filling situation discussed in this paper, there also exists the particle-hole symmetry.

	%(d) and (e) are the intersection of the blue dash line and phase boundary.
	\textit{Ground state.$-$} The ground state of Eq.~(\ref{eq:model}) is always an AFM insulator for any nonzero $J$, captured by exact Hamiltonian
	\begin{equation}
	\hat{H}=\sum_{k\sigma}
	\begin{pmatrix}	
	\begin{smallmatrix}
	\hat{c}_{kA\sigma}^{\dag} & \hat{c}_{kB\sigma}^{\dag}
	\end{smallmatrix}
	\end{pmatrix}
	\begin{pmatrix}	
	\begin{smallmatrix}
	2t'\gamma(k)\sigma+\frac{J\sigma}{4} & -tf(k)\\  %?????
	-tf^{\ast}(k) & -2t'\gamma(k)\sigma-\frac{J\sigma}{4}
	\end{smallmatrix}
	\end{pmatrix}	
	\begin{pmatrix}	
	\begin{smallmatrix}
	\hat{c}_{kA\sigma}\\
	\hat{c}_{kB\sigma}\\
	\end{smallmatrix}
	\end{pmatrix}
	\label{eq:exac_ground}	
	\end{equation}	
	and is inferred by the checkerboard order parameter $\phi_c=\frac{1}{N_s} \sum_{j} (-1)^j\langle q_{j} \rangle$ \cite{kennedy1986itinerant,PhysRevB.87.035128}. Here, $f\left(k\right)=e^{-ik_x}+2e^{\frac{ik_x}{2}}\text{cos}(\frac{\sqrt{3}}{2}k_y)$ and $\gamma(k)=\text{sin}(\sqrt{3}k_y)-2\text{cos}(\frac{3}{2}k_x)\text{sin}(\frac{\sqrt{3}}{2}k_y)$.

		The main results at zero temperature are summarized in Fig.~\ref{fig:0T}. With interplay of topology and magnetism, TIKL model leads to distinct topological and magnetic ground states, \textit{i.e.}, AFMTI, trivial antiferromagnetic insulator (AFMI), TI, Dirac semimetal (DSM) and AFM DSM. These phases have all been predicted in Ref.~\cite{Lieaaw5685,chen2019intrinsic}: (1) paramagnetic (PM) MnBi$_2$Te$_4$ family material is predicted to be DSM/TI and could phenomenologically correspond to the $J=0$, $t'=0$ / $t'\neq0$ case; (2) AFM DSM predicted in MnBi$_2$Te$_4$ corresponds to the boundary of AFMTI-AFMI transition (see Fig.~\ref{fig:0T}); (3) AFMTI and AFMI systems with $J\neq0$ have also been predicted in doped Mn(Sb$_x$Bi$_{(1-x)}$)$_2$Te$_4$ materials \cite{Lieaaw5685}. In the AFMTI phase, different spin flavor contributes opposite Chern number, leading to a TI state with charge Chern number $C_{charge}=0$, spin Chern number $C_{spin}=2$ (invariant $Z_2=1$). This result is consistent with the thickness-dependent magnetic and topological transitions predicted in Ref.~\cite{PhysRevLett.122.107202}.
\begin{figure}[htp!]
	\centering
	\includegraphics[width=3.5in]{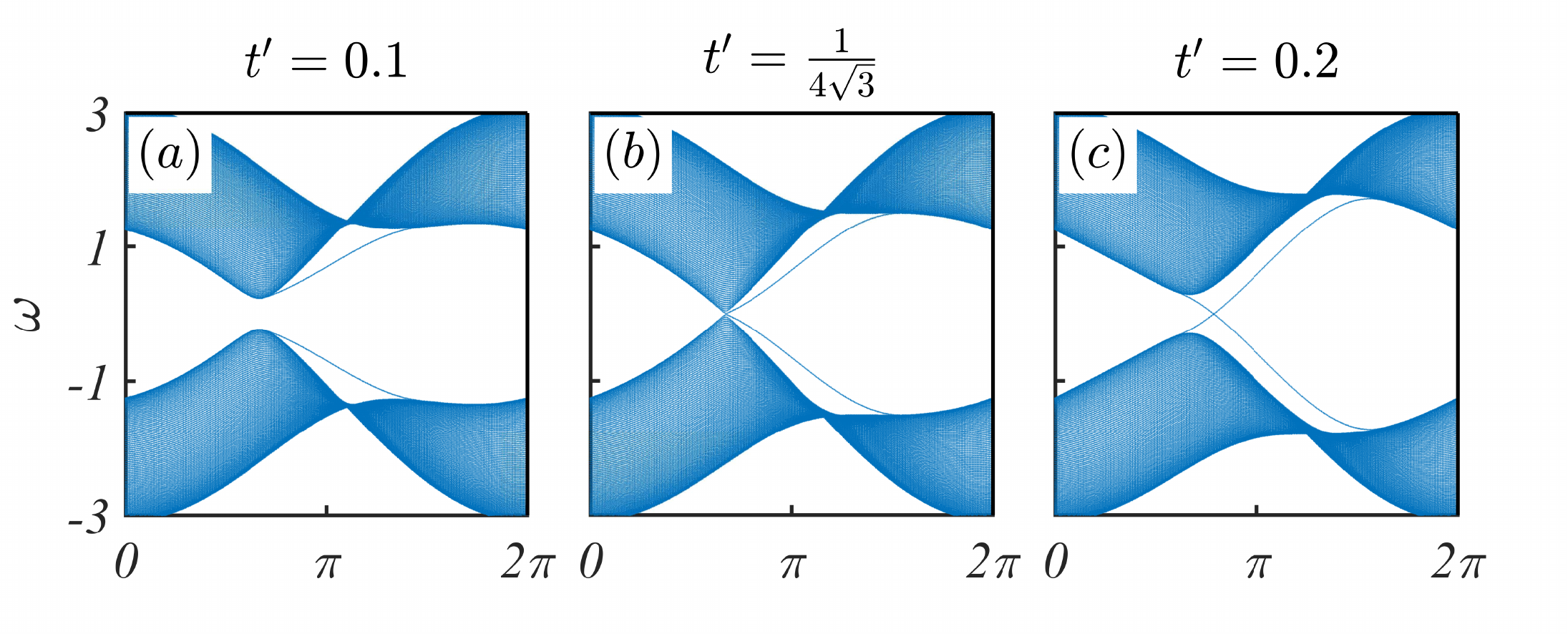}
	\caption{The spectrum of TIKL with the zigzag edges at $J=3$ for (a) AFMI, (b) AFM DSM and (c) AFMTI states. Boundary states sustain to $t'_c$ and disappear with smaller $t'$.}
	\label{fig:Aw2}
\end{figure}		
		Note that in Mn(Sb$_x$Bi$_{(1-x)}$)$_2$Te$_4$ phase diagram \cite{chen2019intrinsic}, there exists a $x$-dependent topological phase transition. When Sb content is increased from $x=0$ to $1$, the Mn(Sb$_x$Bi$_{(1-x)}$)$_2$Te$_4$ crystal would change from AFMTI to AFMI, with band gap closing and reopening at critical $x_c=0.55$ \cite{chen2019intrinsic}. Agreeing with the result of TIKL model, Mn atoms (localized electrons) introduce magnetism, while the Bi-Te/Sb-Te layers (itinerant electrons) control topological states. We report the $x$-dependent topological phase transition could be understood by $t'$-dependent topological phase transition in TIKL model. Therefore, the topological phase transition is fundamentally caused by significant different SOC strength between Bi and Sb, where $\frac{SOC_{Bi}}{SOC_{Sb}}\sim 3.6$ \cite{chen2019intrinsic}. To better understand the topological phase transition, we study the dispersion of $c$-electrons in TIKL. Around Dirac points ($K$) the quasiparticle energy reads $E_{k\sigma}=\pm\sqrt{\frac{9}{4}t^{2}(k-K)^2+(- 3\sqrt{3}t^{{'}}\sigma+\frac{J\sigma}{4})^2}$.  As shown in Fig.~\ref{fig:Aw2}(c), the system in large SOC is a TI with surface states.
		Decreasing SOC ($t'$) leads to a AFMTI-AFMI topological phase transition, which is signaled by the change of $Z_2$ topological number and the vanishing of surface states. %(from $1$ to $0$ together with vanishing of surface states, see Fig.~\ref{fig:Aw2}).
At critical $t'_c=\frac{J}{12\sqrt{3}}$, the band gap $\bigtriangleup=2|E_k|$ first closes and then reopens at the Dirac point (see Fig.~\ref{fig:Ek}), with band inversion disappearing.
%Besides, the location of direct-gap is also shifting from Dirac point $K$ to $M$ with increasing $t'$ (not show here).

		 %In CI, Dirac points $K$ are in the vicinity of the direct-gap (Fig.~\ref{fig:Ek} (d)), which separate with each other upon increasing $t'$ (Fig.~\ref{fig:Ek} (e)). As to trivial AFMI, under a sufficiently large $t'$, the top of energy bands coincide with the high-symmetry points $M$ instead, where topology thoroughly breaks down (Fig.~\ref{fig:Ek} (f)).

	\begin{figure}[htp!]
	\centering
	\includegraphics[width=3.5in]{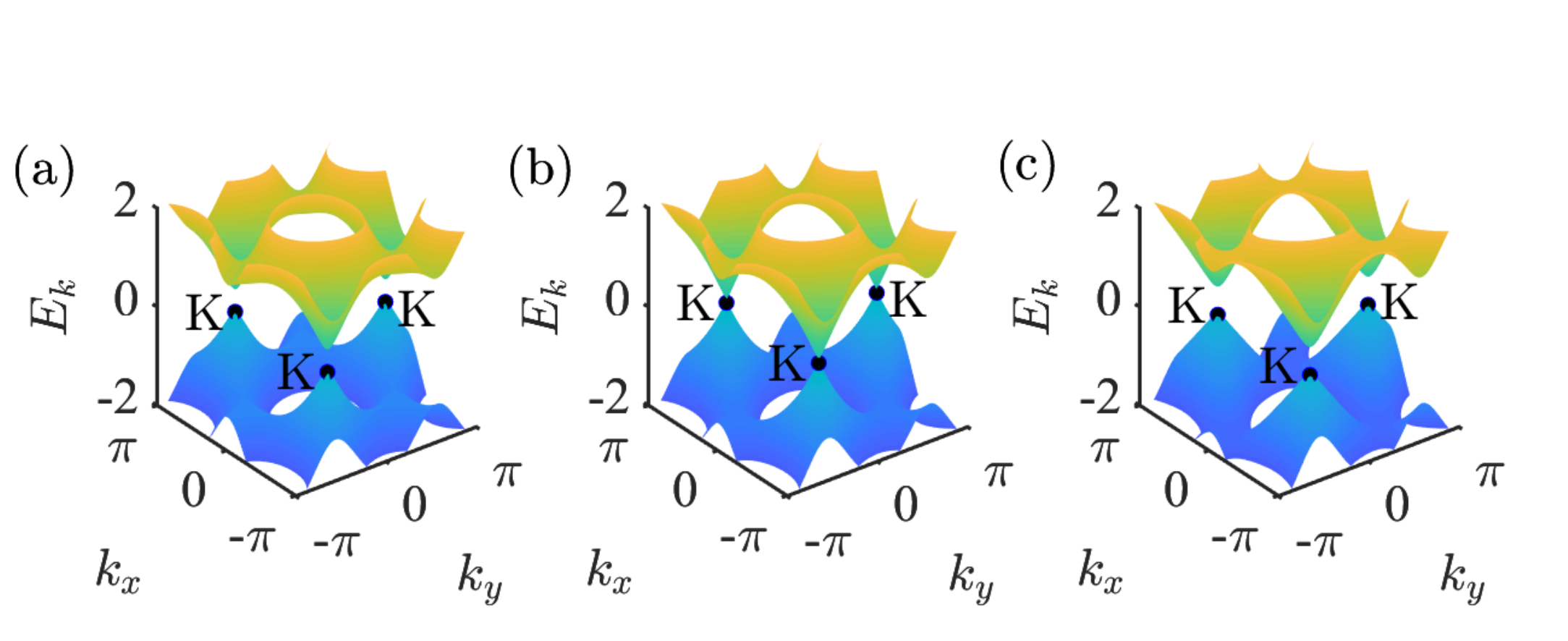}
	\caption{ The $c$-electron's dispersion in the first Brillouin zone at $J=3$ with (a) $t'=0.1$, (b) $t'=\frac{1}{4\sqrt{3}}$  and (c) $t'=0.2$. }
	\label{fig:Ek}
\end{figure}

\begin{figure}[htp!]
	\centering
	\includegraphics[width=3.5in]{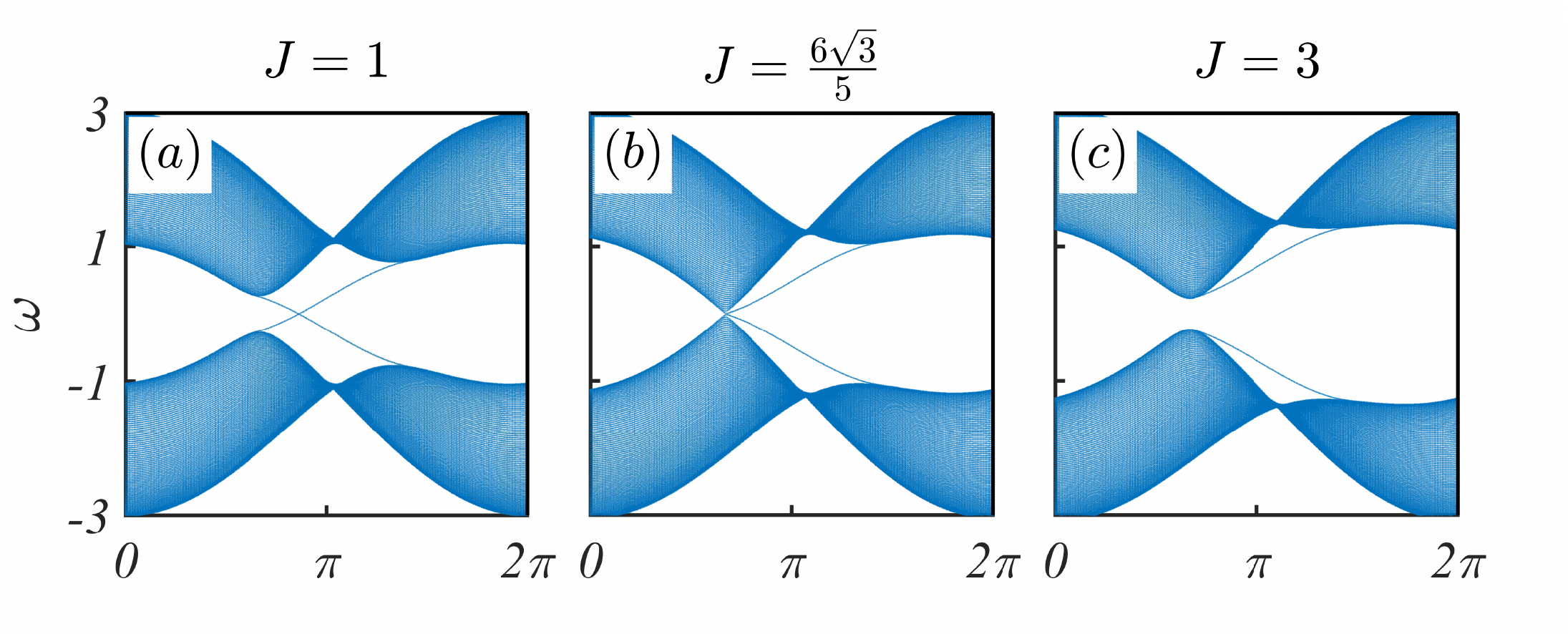}
	\caption{The spectrum of TIKL with the zigzag edges at $t'=0.1$ for (a) AFMTI, (b) AFM DSM and (c) AFMI states. Boundary states sustain to $J_c$ and disappear with larger $J$.}
	\label{fig:Aw1}
\end{figure}

	%(\textit{topological phase transition.})-
 However, apart from SOC, the magnetism is also crucial to topological properties. As shown in Fig.~\ref{fig:0T}, the zero-temperature phase diagram is determined by competition between Kondo coupling and SOC, in which the AFMTI state only exists when $J< J_c$ $(J_c=12\sqrt{3}t')$.  With increasing $J$, the system goes through from a gaped AFMTI
 to a gapless AFM DSM and finally a trivial insulator. On approaching the phase boundary $J_c$, the direct gap $\bigtriangleup_{gap}=2|3\sqrt{3}t'-\frac{J}{4}|$ closes (see Fig.~\ref{fig:Aw1}). Further increasing $J$ makes the gap reopen, corresponding to the thoroughly vanishing of topological properties. Analytical calculation on ground-state energy suggests the topological phase transition belongs to $2D$ Ising universality class \cite{222}. (See details in SM). In layered van der Waals AFM Mn(Sb$_x$Bi$_{(1-x)}$)$_2$Te$_4$ family, the interlayer AFM exchange coupling is quite weak, which has little influence on topology. Instead, the intralayer magnetism do have an effect on topological properties. With Sb content increasing, Mn-Te-Mn bond angle decreases as well as the lattice parameter \cite{PhysRevB.100.104409}. Thus, with reducing Mn-Mn distance the direct in-plane AFM interaction increases and competes with the dominant intralayer FM interaction, leading to reduced single-ion anisotropy and suppressed saturation moment. The stronger intralayer AFM together with larger magnetic frustration in MnSb$_2$Te$_4$ also promotes the $x$-dependent AFMTI-AFMI transition.

%	As found in Ref.~\cite{PhysRevB.74.035109}, excessive SOC ($t>t'_{c}$) also destroys the topology. When NNN hopping dominates, the honeycomb lattice is divided into two interpenetrating triangle sublattices with the lattice constant $\sqrt{3}$ times larger. The topological phase transition is related to the dispersion with different $t'$. In CI, Dirac points $K$ are in the vicinity of the direct-gap (Fig.~\ref{fig:Ek} (d)), which separate with each other upon increasing $t'$ (Fig.~\ref{fig:Ek} (e)). As to trivial AFMI, under a sufficiently large $t'$, the top of energy bands coincide with the high-symmetry points $M$ instead, where topology thoroughly breaks down (Fig.~\ref{fig:Ek} (f)).

	%(\textit{excitation})-

%	\textit{Phenomenological correspondence with MnBi$_2$Te$_4$.-}
	\begin{figure}[htp!]
	\centering
	\includegraphics[width=3in]{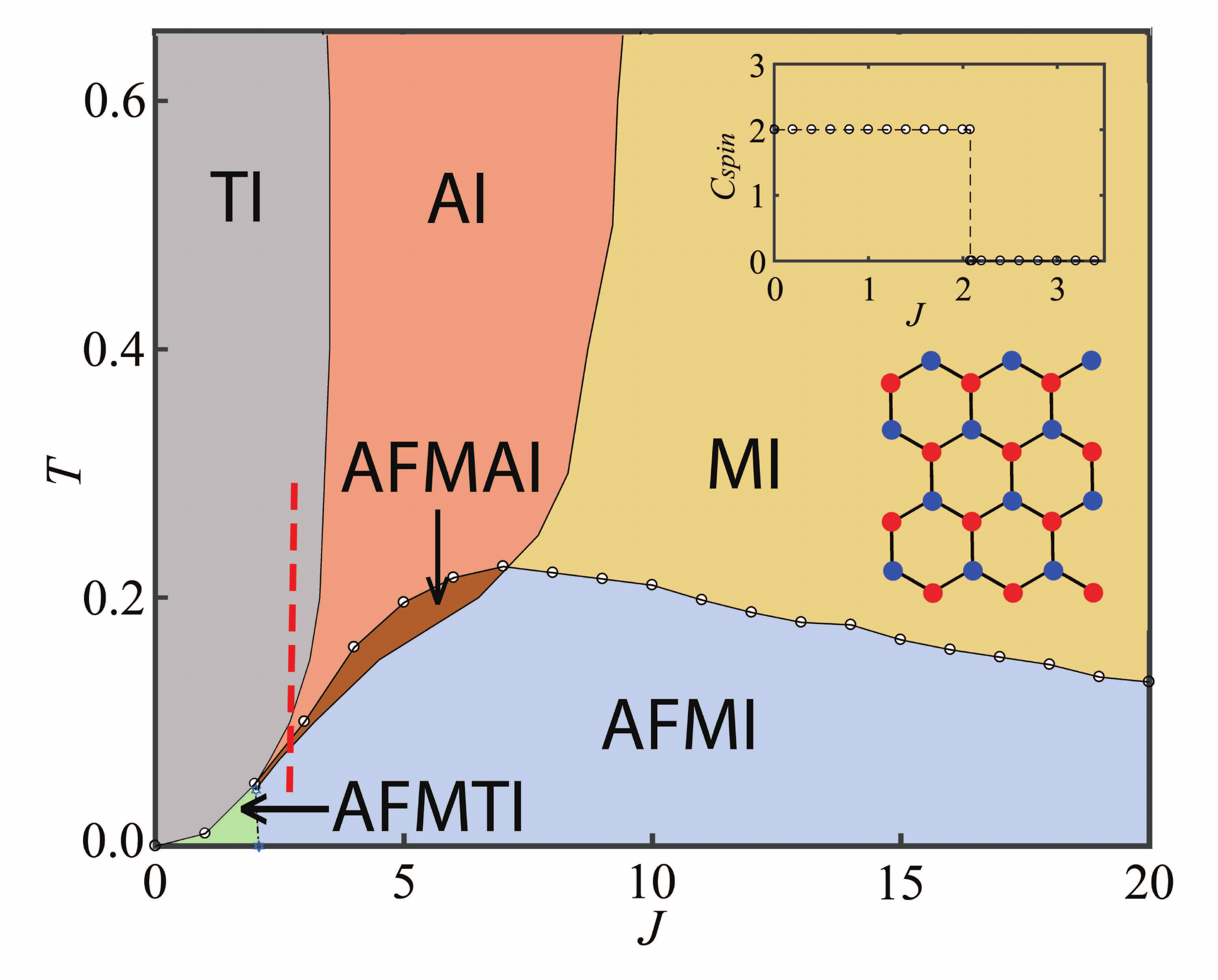}
    \caption{Phase diagram of the particle-hole symmetric TIKL model on two-dimensional in the $T-J$ plane, obtained by lattice Monte Carlo simulations. Low temperature phases: AFMTI at a small $J$ and AFMI at a large $J$. High temperature phases: Anderson insulator (AI) at intermediate $J$ crossing over to a TI at smaller $J$ and a Mott insulator (MI) at larger $J$. At intermediate $T$ and $J$, there exists an antiferromagnetic Anderson insulator (AFMAI) phase. The line with circles indicates the continuous antiferromagnetic-paramagnetic transition. The dashed red line denotes the region that elevated temperature drives topology restoration. Inset: the spin Chern number as a function of $J$ at zero temperature.}
	\label{fig:fig1}
\end{figure}
	
	\textit{Finite-temperature results.}$-$We consider the TIKL with 144 atoms constructed as the inset of Fig.~\ref{fig:fig1} and set $t'=0.1$. We report the main result in the phase diagram (see Fig.~\ref{fig:fig1}). At low temperature, there exists AFMTI at a small $J$ and trivial AFMI at a large $J$. At high temperature, increasing $J$ leads to crossover from PM TI to Anderson insulator (AI) and finally to a Mott insulator (MI). When both temperature $T$ and interaction $J$ are intermediate, an antiferromagnetic Anderson insulator (AFMAI) emerges. Despite the absence of explicit disorder, $f$-electrons at intermediate $J$ instead effectively act as random potential and thus lead to localization in TIKL model \cite{PhysRevLett.117.146601}.

    Although previous studies have revealed the AFM ground state under interplay of topology and Kondo-like interaction \cite{PhysRevB.87.035128}, the resultant AFMTI is limited to the zero temperature situation. In this work we confirm that the AFMTI state do sustain at a considerable finite temperature region. The Kondo interaction introduces a significant gap $\bigtriangleup\sim J$, with which it is promising to realize the high-temperature quantum spin Hall (QSH) effect.

	%(\textit{SDW})-

	%(\textit{SDW transition and crossover.})-
	%At last we focus on properties of transitions.
	In order to confirm the transition properties, we further measure the checkerboard order parameter $\phi_c$, the specific heat $C_v$ and susceptibility $\chi_q$ versus temperature \cite{SupplementalMaterial}. The existence of singularity clearly suggests a phase transition. We make a finite-size scaling analysis of $\phi_c$ under different Kondo coupling, corresponding to the AFMTI-TI, AFMAI-AI and AFMI-MI transitions, respectively, where the AFMTI-TI transition of MnBi$_2$Te$_4$ has been predicted by first-principles density functional theory calculation in Ref.~\cite{Lieaaw5685}.
	Here the Monte Carlo result indicates that the AFM-PM transition is continuous for any nonzero $J$, which belongs to $2D$ Ising universality class.
	At high temperature, there exists merely smooth crossover between TI-AI and AI-MI transitions, where energy density and double occupation of $c$-electrons have linear dependence on $J$ \cite{SupplementalMaterial}.

\begin{figure}[htp!]
	\centering
	\includegraphics[width=0.8\columnwidth]{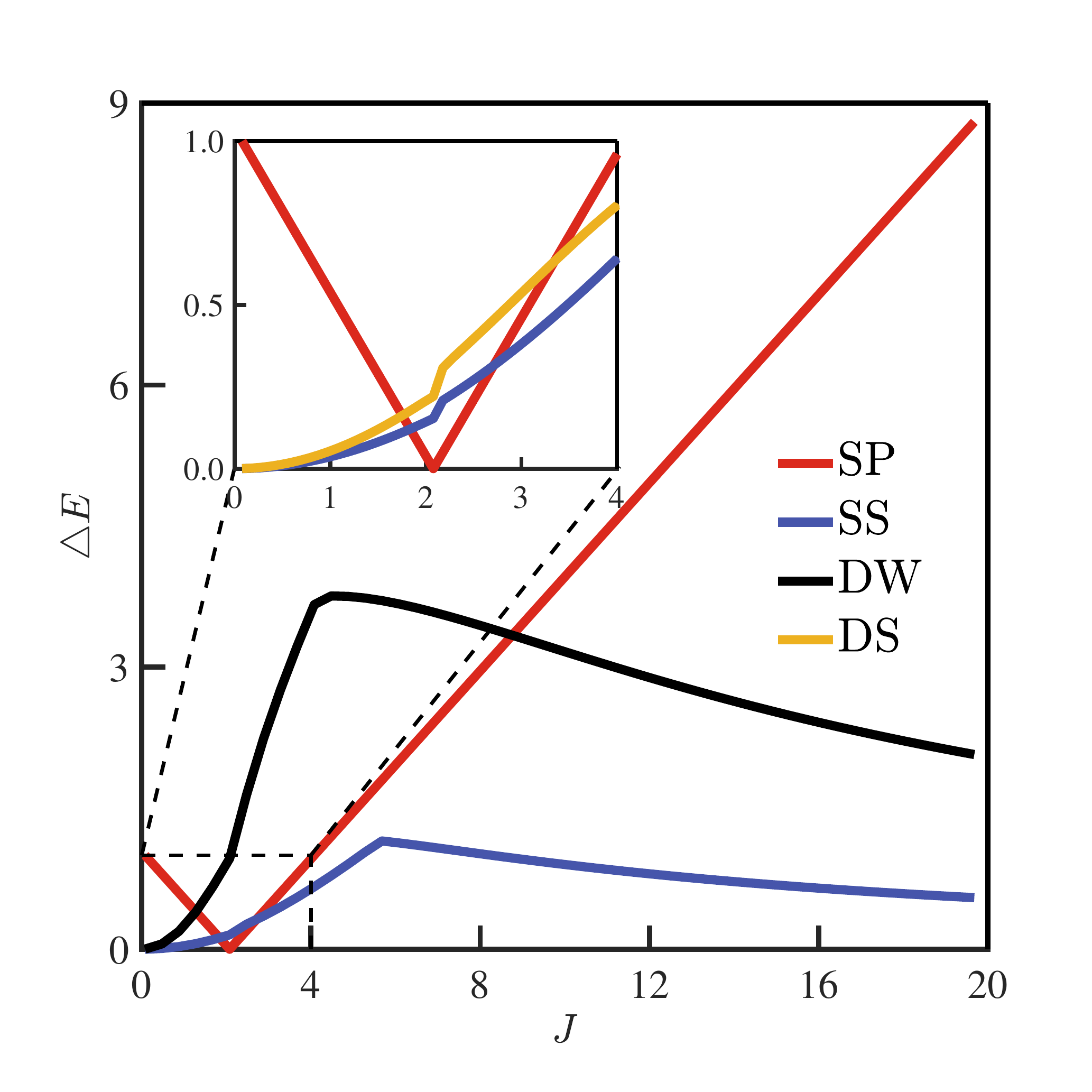}
	\caption{Excited energy above ground state for low-lying excitations: single-particle excitation (SP), single-spin flipping (SS), AFM domain wall (DW) and neighbor double-spin flipping (DS). Inset shows the zoom-in data around topological phase transition.}
	\label{fig:fig_ex}
\end{figure}	

		\textit{low-lying excitation.}$-$Crossing $J_c$ at zero temperature, topology vanishes with gap reopening. When temperature is larger than the energy scale of $c$-electron's excitation (around the red dashed line in Fig.~\ref{fig:fig1}), thermal fluctuation erases the reopened gap and makes the topological properties restore \cite{PhysRevB.93.045138,PhysRevLett.122.126601}. Motivated by this exotic phenomenon, we provide an analysis of some possible low-lying
	excitations. Candidates include: (1) single-particle excitation, (2) single-spin flipping, (3) AFM domain wall and (4) neighbor double-spin flipping. Their excited energy above ground state is depicted in Fig.~\ref{fig:fig_ex}, in which single-spin flipping is the lowest one for most $J$. Except for $J_c$, around which the single-particle excitation is gapless.

	According to energy scale of low-lying excitations under different $J$, the restoration of topology leads to different phases. For convenience, we denote the energy scale of topological gap as $E_{topo}$, and the Kondo-induced one as $E_{Kondo}$. At small $J$ ($J\sim J_c$), where $E_{topo}<T<E_{Kondo}$, erased SOC-induced gap and surviving Kondo-induced gap conspire to a restored TI phase. At large $J$ with $T>E_{topo},T>E_{Kondo}$, both gaps are erased and a topological Anderson insulator (TAI) phase occurs with a metallic density of state. AI is thus divided into TAI and normal AI \cite{111,SupplementalMaterial}. Upon further increasing $J$, the excited energy of single-particle excitation
	increases linearly. When the gap exceeds thermal fluctuation, where $E_{Kondo}<T<E_{topo}$, a normal AI appears with no restoration of topology.

	\textit{Discussion.}$-$%Since many intriguing experimental observations have been reported in mixed valence compound SmB$_{6}$ \cite{neupane2013surface,jiang2013observation,PhysRevB.88.121102}, e.g., smallness of surface velocity and gigantic quantum oscillation signals, our understanding on topological Kondo insulator (TKI) was intensively challenged. These researches triggered many novel ideas like surface Kondo breakdown and Majorana Fermi sea \cite{PhysRevLett.116.046403,PhysRevLett.114.177202,baskaran2015majorana}. An effective method to understand correlation effect in TKI is studying ideal model Hamiltonian, such as topological Kondo lattice (TKL) \cite{PhysRevLett.111.016402,PhysRevB.100.035118,PhysRevLett.104.106408}.
The coexistence of (symmetry-protected) topological order and AFM long-range order has already been predicted in weak coupling regime of topological Kondo insulator (TKI) by a mean-field calculation \cite{PhysRevB.87.035128} and further been investigated by a dynamic mean-field approximation \cite{PhysRevB.93.045138}.
However, the previous theoretical researches about AFMTI state were based on either approximation method or numerical simulation. In this paper, we study the interplay between topology and magnetism with TIKL, which has never been studied in such an exactly analytical way, and the results can also cover many conclusions in standard topological Kondo lattice \cite{PhysRevB.93.045138,PhysRevB.87.035128}. We hope this work could bring some new insights into TKI.

%In TKL, itinerant electrons are modelled by Haldane or Kane-Mele model, so as to generate TKIs when coupled to localized $f$-electron via Kondo interaction.
	%Such antiferromagnet is featured by the coexistence of (symmetry-protected) topological order and antiferromagnetic long-ranged order.
%	Recently, we have revisited the Ising-Kondo lattice (IKL) model\cite{PhysRevB.54.9322,PhysRevB.100.045148}, which serves as the anisotropic limit of
%	Kondo lattice and more importantly its ground states are exactly solvable on bipartite lattice at half-filling, and a disorder-free Anderson localization phenomena occurs at finite temperature.

In topological Kondo lattice, itinerant electrons are modelled by Haldane or Kane-Mele model, so as to generate TKIs when coupled to localized $f$-electrons via Kondo interaction. Here, we choose the Kane-Mele model and thus lead to a solvable TIKL with QSH state. Accordingly, if we model itinerant electrons by Haldane Hamiltonian, the alternative TIKL would instead generate a distinct magnetic topological state, \textit{i.e.}, an AFM Chern insulator with QAH state. (See details in SM).

Most of all, doping Sb in pure MnBi$_2$Te$_4$ will simultaneously introduce the disorder effect, where the chemical disorder effect has been considered in previous researches \cite{PhysRevB.100.104409,chen2019intrinsic}, and thus may produce AI as predicted in Fig.~\ref{fig:fig1}.
Fortunately, the disorder effect will hardly change the global topology, which is protected by a symmetry holding on average \cite{PhysRevLett.109.246605,PhysRevB.89.155424}.
Our work highlights the possibility to realize the TAI in Mn(Sb$_x$Bi$_{(1-x)}$)$_2$Te$_4$ family materials at high temperature. Further careful experiments are needed to investigate the surface states and conductance with precisely regulated doping in Mn(Sb$_x$Bi$_{(1-x)}$)$_2$Te$_4$ materials.

	\textit{Conclusion.}$-$In this work, we utilize the exactly solvable TIKL to phenomenologically describe the Mn(Sb$_x$Bi$_{(1-x)}$)$_2$Te$_4$ family materials. Several previously observed or predicted quantum topological states emerge naturally in TIKL system, including AFM DSM, AFMTI, DSM and TI. It reveals that the TIKL is capable to capture rich physics involved in the intrinsic magnetic TI. In addition, the $x$-dependent topological phase transition could be understood by $t'$-dependent transition in TIKL. With Monte Carlo simulation, we have confirmed the existence of AFMTI at finite temperature, as well as a disorder-free AI. We believe that our findings will offer a promising avenue to realize rich magnetic topological states at high temperature.

This research was supported in part by NSFC under Grant No.~$11704166$, No.~$11834005$, No.~$11874188$, No.~$11674139$.

	%In this work, we have established the existence of topological antiferromagnet and disorder-free AI in a
%	modified Kondo lattice model. The former is the exact ground state if SOC dominates over Kondo coupling and can be restored by thermal fluctuation at finite $T$. At high $T$, localized $f$-electron acts as quenched disorder, leading to the observed Anderson localization. Although novel states of matter like the TKI are the mainstream in heavy fermion community, our work suggests that even the well-understood magnetic phase itself has potential to be topological and can be driven to AI in clear systems.

	\begin{acknowledgements}
		
	\end{acknowledgements}
	
%	\bibliography{reference}

%

\end{document}